\documentclass{article}

\usepackage{arxiv}

\usepackage[utf8]{inputenc} % allow utf-8 input
\usepackage[T1]{fontenc}    % use 8-bit T1 fonts
\usepackage{hyperref}       % hyperlinks
\usepackage{url}            % simple URL typesetting
\usepackage{booktabs}       % professional-quality tables
\usepackage{amsfonts}       % blackboard math symbols
\usepackage{nicefrac}       % compact symbols for 1/2, etc.
\usepackage{microtype}      % microtypography
\usepackage{graphicx}
\usepackage{natbib}
\usepackage{doi}
\usepackage{listings}
\graphicspath{ {./graphics/} }

\title{WaCoDiS: Automated Earth Observation Data Processing within an Event-Driven Architecture for Water Monitoring}

\author{
    Sebastian Drost\\
    Bochum University of Applied Sciences\\
	Department of Geodesy\\
	Am Hochschulcampus 1 \\
	D-44801 Bochum \\
	\texttt{sebastian.drost@hs-bochum.de} \\
	\And
	Arne Vogt \\
	52°North Spatial Information Research GmbH\\
	Martin-Luther-King-Weg 24\\
    D-48155 Münster, Germany\\
	\And
	Christian Danowski-Buhren \\
	Bochum University of Applied Sciences\\
	Department of Geodesy\\
	Am Hochschulcampus 1 \\
	D-44801 Bochum \\
	\And
	Simon Jirka \\
	52°North Spatial Information Research GmbH\\
	Martin-Luther-King-Weg 24\\
    D-48155 Münster, Germany\\
    \And
	Verena Kirstein\\
	Wupperverband\\
	Untere Lichtenplatzerstraße 100\\
    D-42289, Wuppertal\\
    \And
	Kian Pakzad\\
	EFTAS Fernerkundung Technologietransfer GmbH\\
	Oststraße 2-18\\
	D-48145, Münster\\
    \And
	Matthes Rieke \\
	52°North Spatial Information Research GmbH\\
	Martin-Luther-King-Weg 24\\
    D-48155 Münster, Germany\\
}

\date{}

\renewcommand{\shorttitle}{WaCoDiS: Automated Earth Observation Data Processing}

\hypersetup{
pdftitle={WaCoDiS: Automated Earth Observation Data Processing within an Event-Driven Architecture for Water Monitoring},
pdfauthor={Sebastian~Drost, Arne~Vogt, Christian~Danowski-Buhren, Simon~Jirka, Verena~Kirstein, Kian~Pakzad, Matthes~Rieke},
pdfkeywords={Water Monitoring, Earth Observation, Copernicus Satellite Data, Event-Driven Architectures, Spatio-Temporal Data},
}

\begin{document}
\maketitle

\begin{abstract}
Climate changes and the ongoing intensification of agriculture effect in increased material inputs in watercourses and dams. To ensure an efficient and environmentally friendly water resource management, water management associations need means for efficient water monitoring as well as novel strategies to reduce the pollution of surface and ground water. Traditionally, water management associations operate large sensor networks to suffice their needs for hydrological and meteorological measurement data to monitor and model physical processes within the catchments they are responsible for. Implementing a comprehensive monitoring system often suffers from sparse coverage of in-situ data. Due to the evolvement of the Copernicus satellite platforms, the broader availability of satellite data provides a great potential for deriving valuable, complementary information from Earth Observation data, that contributes to a detailed understanding of hydrological processes. Although the number of satellite data platforms that provide online processing environments is growing, it is still a big challenge to integrate those platforms into traditional workflows of users from environmental domains such as hydrology. Thus, in this paper, we introduce a software architecture to facilitate the generation of Earth Observation information targeted towards hydrology. The presented WaCoDiS System comprises several microservices as well standardized interfaces that enable a platform-independent processing of satellite data. First, we discuss the contribution of Earth Observation data to water monitoring and derive several challenges regarding the facilitation of satellite data processing. We then describe our system design with a brief overview about the different system components which form an automated processing pipeline. The suitability of our system is proven as part of a pre-operational deployment for a German water management association. In addition, we demonstrate how our system is capable of integrating satellite data platforms, using the Copernicus Data and Exploitation Platform - Deutschland (CODE-DE) as a reference example.
\end{abstract}

% keywords can be removed

\section{Introduction}
In recent years, ensuring the quality of drinking water has become one of the most challenging tasks for water management associations. Especially, the intensification of agriculture has impacts on the water quality of ground and surface water. The enhanced efficiency in agricultural production often comes with increased fertilization which leads to nutrient leaching and a higher level of nitrate concentration in groundwater \citep{parris_impact_2011}. In addition, due to climate changes, the number of extreme weather events, such as regional heavy rainfall or long-term droughts, increased and amplified the problem \citep{klages_impact_2020}. Hence, water management associations more frequently must deal with pollutant inputs into ground and surface water. To ensure an environmentally friendly water resource management, there is a great need for improved water monitoring programs as well as strategies to reduce the pollution of drinking water \citep{kirschke_agricultural_2019}.

Hydrological models, that are capable for estimating the environmental impact of land management practices, contribute to the implementation of a comprehensive water monitoring strategy. However, such models often suffer from large uncertainty due to the sparse availability of highly accurate in-situ data \citep{seibert_gauging_2009}. It could be shown that the understanding of hydrological processes, such as soil moisture determination, evapotranspiration estimation, or streamflow prediction benefit from the exploitation of satellite data \citep{hulsman_learning_2021}. At the same time, the evolvement of several Copernicus satellite data platforms, such as Copernicus Data and Exploitation Platform - Deutschland (CODE-DE), the Copernicus Open Access Hub and the Copernicus Data and Information Access Services (DIAS), which facilitate the discovery and access of suitable Copernicus satellite data, lead to a frequent emergence of advanced environmental monitoring applications \citep{gomes_overview_2020}.

However, generating valuable information from Earth Observation data (EO data) as well as utilizing the platforms’ processing capabilities require in-depth knowledge in several topics such as remote sensing, data science and cloud computing. Thus, applications and users often do not exploit the full potential of those platforms which results in less efficient processing workflows. To overcome these issues, the research project WaCoDiS (Copernicus-based services for monitoring material inputs in watercourses and dams) addressed several requirements of users from the hydrology regarding an efficient processing of large earth observation data, to improve water monitoring. As part of the project, a modular and extensible microservice software architecture has been developed that connects to Copernicus satellite data platforms and implements an automatic processing workflow for generating Earth Observation products. Thus, the present paper introduces an approach for efficient online processing of large Earth Observation data in cloud environments relying on interoperable Open Geospatial Consortium (OGC) interfaces. It covers different architectural aspects addressing the need for an event-driven communication – i.a. publish/subscribe patterns and messaging protocols – as well as novel strategies for injecting ready-to-use products into existing information infrastructures.

\section{Contribution of Earth Observation data to water monitoring}
The research project WaCoDiS aimed to improve typical water monitoring tasks, considering the 813 km² large catchment area of the Wupper river in Germany as research area. As responsible water management association for this region, the Wupperverband (WV) deals with several water management tasks, including sewage treatment as well as water quantity control. To fulfill its responsibilities, the WV continuously monitors water bodies and water facilities, such as dams and sewage treatment plants, through automatic and manual measurements. This includes continuously measured water level data as well as water quality parameters such as pH, conductivity, temperature and oxygen.

The spatial coverage of these data sets is very focused since they solely cover certain critical infrastructures (EU-Directive 2008/114/EG), which justify the rather high in-situ data collection costs (e.g., dedicated instrumentation, maintenance). At the same time, the spatial and temporal coverage of other areas beyond these focused locations, is rather low. Consequently, for important challenges such as quantifying material inputs into surface water bodies, the existing in-situ data is not yet sufficient.

To evaluate the opportunities of using EO data as a complementary data source in water monitoring, a systematic requirements analysis was conducted within the Wupperverband. The results have shown that remote sensing and derived products are a promising and cost-efficient method for an enhanced water quality monitoring. With the use of remote sensing data and their derived thematic products, an overall picture of the monitored area at a high spatial resolution and coverage can be achieved. This helps closing spatial gaps in punctual measurements such as of soil moisture \citep{huang_detecting_2018}. Due to the increased spatial spreading of hydrological information received by remote sensing, several facilities can be monitored for similar changes at the same time or based on historic proceedings. The health of water bodies depends on impacts such as material inputs due to weather conditions, land cultivation or soil condition, which are currently surveyed only based on specific incidents (e.g., heavy rain events, droughts) or selectively. With the use of remote sensing data inconsistencies within the monitored water quality parameters can be investigated to identify potential causes and justify continuous monitoring at specific places \citep{dornhofer_remote_2016}.

In addition, information derived from EO data can be integrated into hydrological models used for geolocating and quantifying sediment and material inputs into surface water bodies. Recent studies have shown, that EO data products are capable of optimizing hydrological models, that often suffer from the limited coverage of highly accurate in-situ data \citep{lopez_lopez_calibration_2017}. The distribution of land cover in the catchment area supports the calculation of sediment and material inputs into rivers, the prospective amount of rainwater entry as well as the vitality of forests and vegetation. Monitoring the growing on agricultural areas within a catchment area, enables insights into the nitrate pollution of surface waters \citep{preidl_introducing_2020}. Land cover and land use are therefore significant indicators for environmental impacts caused by human actions. However, the temporal resolution of publicly available land-cover products such as Corine Land Cover varies between three to six years and is mostly rough, meaning that only general categories are defined. To improve water monitoring, a more specific land cover and land use classification with a higher temporal resolution is indispensable to react on changes in land use in a timely manner. For instance, a fine-grained distinction between different types of agricultural areas contributes to analyzing the impact of land-use changes in watershed discharge.

\section{The potential of Earth Observation data analytics}
The combination of in-situ and remote sensing methods enables a time- and cost-efficient large-scale water monitoring, which derives accurate estimates regarding water quality in a small-scale area \citep{govender_review_2009}. In particular, remote sensing methods are able to collect data for rather large areas at once. Especially satellite imagery provides frequent observations of larger parts of the surface of the earth. Programs such as the European Copernicus initiative\footnote{\url{https://www.copernicus.eu/}} are now offering valuable data without license cost.

In discussion with stakeholders from practice, especially the different departments of the Wupperverband and further water associations, a set of potential information products have been identified which might be derived from Earth Observation data. These information products serve as test cases for the use of EO data that has been addressed in WaCoDiS and the automated processing approach presented in this paper:
\begin{itemize}
	\item Detailed land use classification: Describes the physical or biological coverage of the Earth's surface. It can be linked to (human) land use and associated pollutant emissions. Moreover, it is the basis for erosion modeling, based on \citet{breiman_random_2001}.
	\item Intra-annual land use classification: In addition to the detailed land use classification this product enables an intra-annual classification for agricultural areas..
	\item Leaf Area Index (LAI): Quantifies the one-sided green leaf area per unit ground surface area, thereby characterizing plant canopies. It is an essential parameter for the estimation of evapotranspiration, based on \citet{weiss_s2toolbox_2016}.
	\item Chlorophyll-a concentration in water bodies: Chlorophyll-a and total suspended matter data is obtained from water surfaces to monitor and model water quality on a large scale, based on \citet{brockmann_evolution_2016}.
	\item Surface temperature (water and ground) based on \citet{yu_land_2014}, \citet{sobrino_land_2008} and \citet{barsi_atmospheric_2003}.
	\item Urban Soil Sealing: Provides detailed information on the permeability of settlement areas, which impact surface runoff and substance transport as well as microclimatic processes based on \citet{kaspersen_using_2015}.
	\item Analysis of vitality changes of vegetation: The vitality changes of vegetation, e.g., for forest areas, can give an important hint of changes in the soil or indicate bark-beetle infestations.
\end{itemize}

\section{Related work on Earth Observation data processing}
For designing an efficient approach to enable (event-driven) processing of Earth Observation data, there are several trends from information technology that need to be considered. This comprises for example infrastructure aspects and cloud technologies, event-based communication and processing flows as well as interoperability standards.

The extremely high volume of data generated by satellites leads to new challenges in data processing (the Copernicus Sentinel Data Access System publishes around 15 TB of data per day \citep{european_space_agency_copernicus_2019}. In the view of this immense amount of data, new workflows are needed to explore the available data. Also, the volume of input scenes (the number of available scenes and the size of each individual scene) is increasing so that conventional workflows such as downloading data locally for any processing become inefficient. Thus, cloud computing concepts are used to provide dedicated platforms for discovering and processing Earth Observation data \cite{gomes_overview_2020}. In the Copernicus ecosystem, especially the Data and Information Access Services (DIAS) platforms such as WEkEO\footnote{\url{https://www.wekeo.eu/}}  and Mundi Web Services\footnote{\url{https://mundiwebservices.com/}} need to be mentioned. These European approaches are further complemented by nationally operated platforms such as CODE-DE\footnote{\url{https://code-de.org/}}.

Complementary to the new opportunities of cloud-based data management, also event-based communication flows offer a great potential for the design of our system approach. This allows to actively send information and trigger workflows as soon as new observation data sets become available. On the one hand, the topic of event-based workflows has been taken up by international standardization organizations such as the Open Geospatial Consortium with the OGC Publish/Subscribe Interface Standard \citep{braeckel_ogc_2016}. On the other hand, more and more technologies and protocols are available to implement event-based workflows in practice. An important example is the Advanced Message Queuing Protocol (AMQP) \citep{oasis_oasis_2012} with a range of available open-source implementations such as RabbitMQ\footnote{\url{https://www.rabbitmq.com/}}. Also, frameworks for stream-based data processing (e.g., Apache Kafka\footnote{\url{https://kafka.apache.org/}}) help to enable event-based data analysis workflows.

The third important technical building block that helps enabling a modular system design are interoperability standards for encapsulating data processing components. Here, especially the developments of the OGC with regards to the OGC Web Processing Service (WPS) \citep{muller_ogc_2015} and its advancement into the OGC API for Processes \citep{pros_ogc_2020} provide a valuable means. Relying on these standards, it is possible to encapsulate any kind of process into an exchangeable module that exposes a common interface.

\section{Facilitating Big Earth Observation data processing}
Due to the advent of cloud computing technologies, web-based processing and analysis workflows more and more replace download-driven approaches. The increasing amount of satellite image data leads to higher hardware requirements to be able to generate valuable information from EO data. In consequence, traditional processing workflows, which are based on downloading and local processing of satellite image data, became quite inefficient.

Platforms for Big Earth Observation data already addresses different challenges regarding storing, sharing and processing of large geospatial datasets. Several providers rely on production-ready cloud solutions which uses efficient storage mechanism for distributed systems and provide computing environments that enable a scalable server-side processing with user applications \citep{gomes_overview_2020}.

However, the paradigm shift in processing Big EO Data comes with new challenges. Regarding to \citet{sudmanns_big_2020} the main future tasks in the EO community comprise:
\begin{itemize}
	\item sharing expert knowledge, in the form of EO processing workflows, algorithms and analysis information for non-expert users,
	\item facilitating the accessibility and usability of cloud computing platforms for non-programmers,
	\item increasing the interoperability and portability of EO processing services across different cloud providers,
	\item support for reproducibility of processing results to increase the trust in EO based information.
\end{itemize}
The following sections will address those challenges by identifying and deriving specific architecture requirements for a distributed software system that aims to fill the gap between all related parties:
\begin{itemize}
	\item users from environmental domains (i.e. hydrology within the context of WaCoDiS), that are interested in certain Earth information, but have only little knowledge in EO data processing,
	\item the EO community, that has in-depth knowledge in EO data processing and the underlying algorithms,
	\item and the cloud platforms that provide efficient technical solutions for common EO related tasks.
\end{itemize}

\subsection{Automated processing pipeline}
\label{Automated processing pipeline}

The application of EO data and analysis tools for target-oriented monitoring and evaluation is conducted in several successive steps. In general, we identified the following actions for an EO data processing pipeline:
\begin{enumerate}
	\item \textbf{Process Definition:} Specification of the domain task, the area of interest and applicable processing tools.
	\item \textbf{Data Discovery:} Identification of relevant satellite data products for the target area of interest by observing suitable remote sensing data platforms.
	\item \textbf{Data Retrieval:} Query of required remote sensing datasets (and maybe other required resources).
	\item \textbf{Data Processing:} Preparation and execution of dedicated EO processing tool(s) analyzing the previously gathered data and deriving results in form of higher-quality EO products.
	\item \textbf{Result Collection and Distribution:} Acquisition of processing results/products and distribution within information infrastructures.
	\item \textbf{Result Inspection:} Subsequent treatment and analysis of the computed results.
	\item \textbf{Workflow and Result Evaluation:} Technical assessment of both, the whole processing workflow and resulting EO products performed by specialists.
\end{enumerate}

Hereby, the first processing item (Process Definition) defines the foundation for the remaining actions and must be created by a domain expert once. In particular, the remaining steps may be repeated periodically i.e., in interval-based monitoring scenarios, or be triggered event-based as a result of a certain incident (e.g., heavy rainfall within the context of hydrology).

While manual execution of this looped processing chain requires deep knowledge in several domain-related and technological fields, recent developments in distributed (cloud) computing allow for vast automation of the outlined EO processing pipeline. Remarkably, an automatable and replicable implementation of the key tasks Data Discovery, Data Retrieval, Data Processing, and Result Distribution for arbitrary user defined EO domain scenarios may not only reduce the need for manual user interaction but, especially, increase processing stability and efficiency.

\subsection{Interoperable processing of Earth Observation data}
Several cloud platforms for Earth Observation data provide a scalable computing environment for online processing but differentiate in the level a user can interact with those systems \citep{sudmanns_semantic_2018}. The lowest level of interaction is provided by platforms, that restrict the user to only apply existing tools or processing workflows to selected datasets, whereas platforms, that allow the deployment and execution of algorithms developed by users, are most flexible. Commonly, open-source tools for processing Earth Observation data such as the Sentinel Toolboxes\footnote{\url{https://sentinel.esa.int/web/sentinel/toolboxes}} or the Orfeo ToolBox\footnote{\url{https://www.orfeo-toolbox.org/}} as well as programming libraries for raster data like GDAL\footnote{\url{https://gdal.org/}} form the backbone of the platforms’ processing capabilities.

However, the flexibility to run arbitrary tools on those platforms comes with different challenges regarding the interoperability and usability of Earth Observation data processing within the proposed system. In order to integrate different EO processing services for water monitoring tasks in automated workflows, some essential requirements are derived. The following key aspects take into account the differences in handling varying cloud platforms’ APIs and different processing tools:
\begin{enumerate}
    \item To exploit the potential of existing EO tools as well as custom algorithms but ensure its automatic execution, a common web-based interface should encapsulate EO processing services.
    \item The integration of EO processing services in automated workflows should be possible, regardless of the cloud platforms they are deployed on. Thus, process executions must be agnostic to different processing environments (different cloud computing platforms as well as self-hosted environments are possible).
    \item To support the use of EO processing services in different client applications, the use of standardized technologies should guarantee its interoperability.
\end{enumerate}

\subsection{Sharing Earth Observation information}
Analysis results and information products generated from EO data often build the basis for extended processing chains within certain expert fields. Domain expert users integrate EO products into own applications or combine them with heterogeneous spatio-temporal datasets to gain further insights into domain related problems. Thus, information products must be shared in a way that allows their reliable utilization for domain tasks. This requirement not only relates to the provision of EO products within the users' environment but also to the reproducibility of the shared information \cite{goswein_data_2019}. 

Regarding to the use-cases addressed in the WaCoDiS research project, EO products should become available to support domain experts of a regional water management association in handling water management related problems. The domain users operate on different spatio-temporal datasets that are provided within a Spatial Data Infrastructure which comprises hydrological in-situ measurements and business data as well as different geospatial data services. This leads to the requirement, that newly generated EO data products, which in general are raster datasets, must be ingested into the existing infrastructure.

Furthermore, the data ingestion must be triggered as soon as newly generated products have become available to suffice event-driven data flows. This also addresses the requirement of a completely automated processing workflow. In addition, for future use cases it is conceivable that EO information products should be shared with other organizations or the general public. Therefore, the data ingestion approach should be designed to serve various infrastructures, which are not known in beforehand. In short terms, the proposed system must provide the flexibility to ingest EO data products into an arbitrary number of backends, which are placed in existing infrastructures and connect to the system in a plug and play manner using publish/subscribe patterns.

\section{Design of an event-driven microservice architecture}
Concerning the architectural requirements of an interoperable and automated processing environment for arbitrary EO data, the WaCoDiS platform enables domain users to create repeatable processing jobs through the specification of required input data for a certain spatial and temporal coverage as well as supported EO computation tools. The WaCoDiS platform facilitates an event-driven microservice approach to automatically trigger and execute the EO processing pipeline sketched in section \ref{Automated processing pipeline}. In detail, WaCoDiS discovers and retrieves required data from available EO data providers, schedules the associated process execution and automatically ingests resulting EO products into existing data infrastructures.

For this purpose, current state-of-the-art web technologies within the scope of distributed cloud computing are combined with international web standards in order to implement a highly scalable, modular and event-driven processing platform within the EO context. The following sections presents the approach in detail. 

\subsection{System components for processing tasks}
\label{System components for processing tasks}

Based on the information of a WaCoDiS process description (\textit{WacodisJobDefinition}), the WaCoDiS platform automatically performs the process preparation and execution according to the EO processing pipeline outlined in section \ref{Automated processing pipeline}. Such a \textit{WacodisJobDefinition} covers certain information describing all relevant aspects for executing a process, such as the execution interval in form of a crontab pattern, the temporal and spatial coverage as well as the desired EO process and required input datasets (Listing \ref{lst:wacodisjobdefinition}). 

\begin{lstlisting}[caption={Exemplary WacodisJobDefinition, describing the periodical execution of a land cover classification every first day in a month, using Sentinel-2 satellite data that cover the last 14 days (partial details are omitted for brevity)}, captionpos=b, label={lst:wacodisjobdefinition}]
{
  "id": "851956cb-0975-407c-bada-a08247f13c5c",
  "name": "Land cover classification catchment area",
  "description": "Monthly executed land cover classification.",
  "created": "2020-07-07T12:03:26.006Z",
  "lastFinishedExecution": "2020-07-07T14:00:00.000Z",
  "status": "waiting",
  "execution": {
    "pattern": "0 0 1 * *"
  },
  "temporalCoverage": {
    "duration": "P14D"
  },
  "areaOfInterest": {
    "extent": [6.931,50.985,7.607,51.319]
  },
  "processingTool": "de.hsbo.wacodis.land_cover_classification",
  "productCollection": "land-cover-classification",
  "inputs": [
     {
       "sourceType": "CopernicusSubsetDefinition",
       "identifier": "opticalImage",
       "maximumCloudCoverage": 20.0,
       "satellite": "sentinel-2"
     }
   ]
}
\end{lstlisting}

\begin{figure}[htp]
    \centering
    \includegraphics[width=12cm]{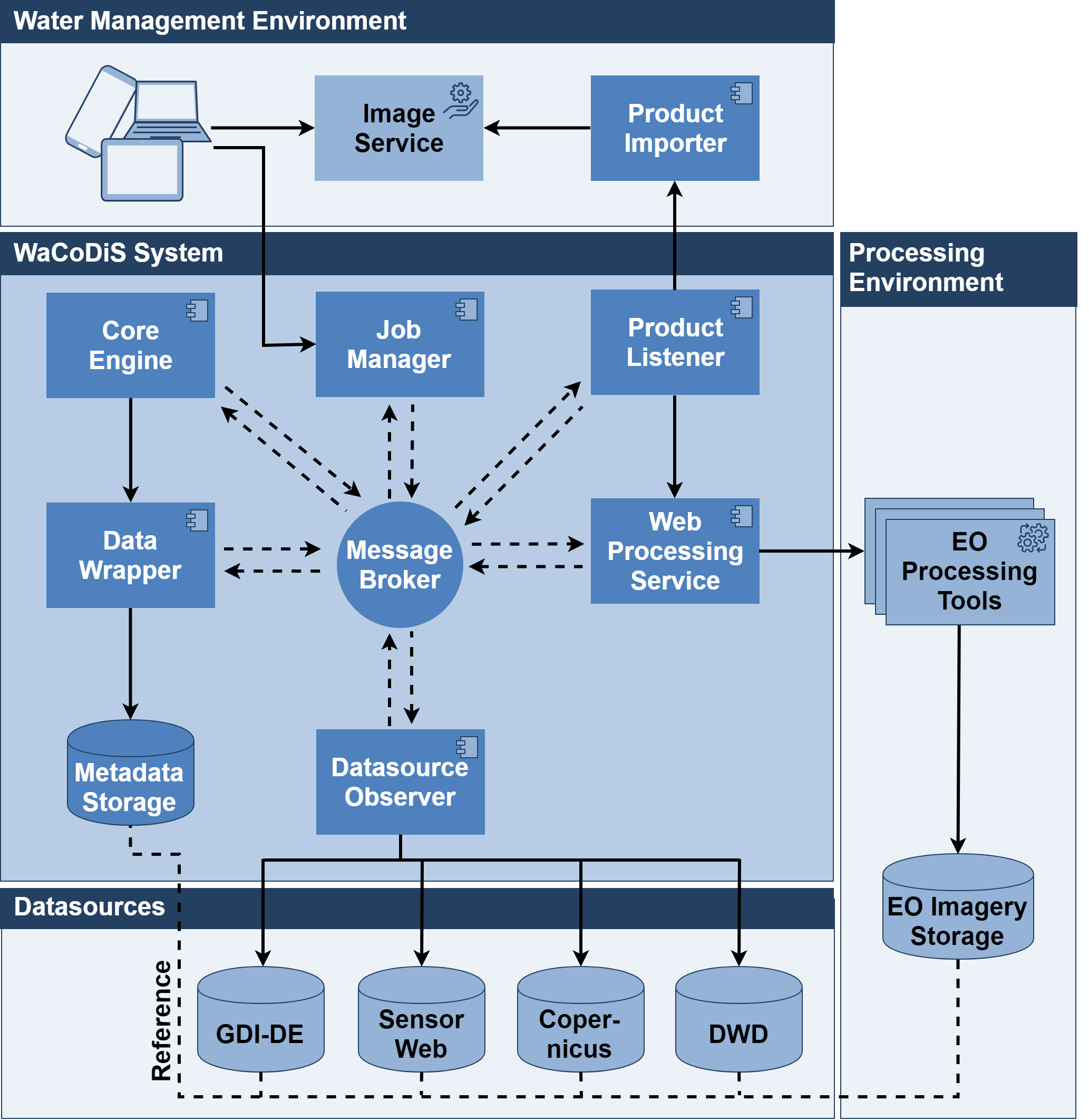}
    \caption{Architectural overview of the WaCoDiS System components (dark blue coloration emphasizes novel components building the WaCoDiS System; light blue coloration indicates legacy tools, services and data sources}
    \label{fig:wacodis_architecture}
\end{figure}

Once provided to the WaCoDiS System, the process description will be handled by different lightweight components, which are embedded in a microservice-oriented architecture (Figure  \ref{fig:wacodis_architecture}). Each of the subsequently described components focuses on a single dedicated task which is part of the processing pipeline.

\begin{description}
   \item[Job Manager:] The Job Manager stores \textit{WacodisJobDefinitions} and serves as the entry point for system users to submit new processing tasks to the WaCoDiS System.
    \item[Datasource Observer:] Each \textit{WacodisJobDefinition} comprises a list of required process input datasets (e.g., satellite scenes, weather data and other data types in form of a \textit{WacodisSubsetDefinition}). To achieve full automation of data retrieval, the Datasource Observer component searches for matching datasets. Considering the temporal coverage definition, the Datasource Observer observes the respective data platform, queries metadata about available datasets and submits this metadata (via the message broker) to the WaCoDiS Data Wrapper component. The compiled metadata comprises domain-specific information and details on where and how to download the actual dataset.
    \item[Data Wrapper:] Domain-specific metadata including availability and accessibility information for relevant processing input datasets is managed by a REST-based Data Wrapper component. The component is responsible for creating download references to physical datasets. Therefore, it considers different dataset related conditions. For example, if Sentinel-2 satellite data is requested for performing a processing task, the component generates a download link to a suitable dataset on the CODE-DE platform, taking into account temporal and spatial requirements as well as cloud coverage limitations and the processing level. 
    \item[Core Engine:] The Core Engine acts as the scheduling component of the whole system. It considers the repetition and trigger rules for registered processing jobs in order to notify the respective components to initiate data querying and process execution.
    \item[EO Processing Tools:] WaCoDiS comprises an expandable set of processing tools, each implementing an executable EO processing algorithm. The actual processing algorithms operate within the processing environment of satellite data platforms or within a self-hosted environment and generate the Earth Observation products.
    \item[Web Processing Service:] To guarantee a unified and standardized way for processing EO data, the OGC Web Processing Service (WPS) standard is used within the WaCoDiS System. The WPS server resolves the references, which are passed as process inputs, by fetching or downloading the data, applies basic preprocessing routines and triggers the EO Processing Tool for generating the target EO product
    \item[Product Listener:] Once a desired EO product is successfully computed, a Product Listener fetches the results and ingests it into a target information data infrastructure.
    \item[Message Broker:] The event-driven microservice architecture of WaCoDiS relies on asynchronous message exchanges via a central message broker, thus enabling a high scalability and flexible deployment of the individual system components on multiple hosts. According to the Publish/Subscribe pattern, WaCoDiS components send messages containing certain process related information to defined channels of the message broker, while subscribing components listen for those messages, they need for fulfilling their tasks.
\end{description}

\subsection{WPS encapsulation of EO data analytics services}
The OGC WPS specification is a well-established interface standard for web-based geospatial processing \citep{muller_ogc_2015}. Its applicability for Big EO data processing within cloud environments has been proven in recent studies \citep{chen_cloud_2012, zhang_integrating_2017} and was one of the topics that were addressed as part of \citep{simonis_standardized_2018}. As an essential outcome, the utilization of virtualization and containerization technologies in combination with the WPS interface enables a standardized and scalable execution of any kind of application within cloud environments. Consequently, the EO data processing approach in WaCoDiS heavily relies on a WPS server that wraps pre-deployed processing scripts as well as custom containerized applications to ensure interoperability.

\begin{figure}[t]
    \centering
    \includegraphics[width=12cm]{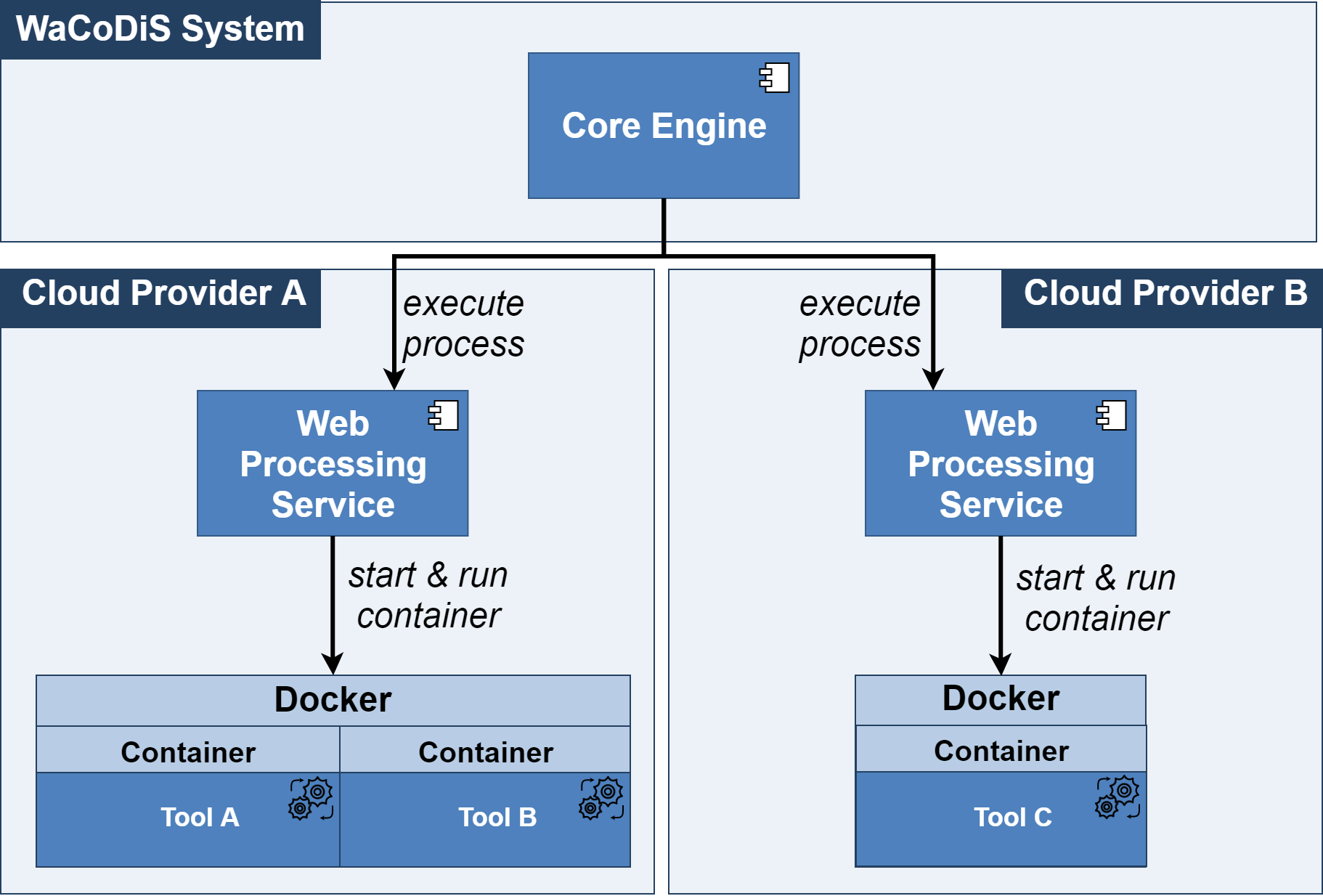}
    \caption{WPS encapsulation of EO tools across different cloud platforms to support interoperability}
    \label{fig:wps_processing}
\end{figure}

The WPS interface allows the execution of any kind of process in a standardized way, which provides the flexibility to comprise custom developed algorithms directly within a WPS implementation as well as to trigger the execution of external tools and scripts. In addition, the WPS server provides detailed descriptions of supported processes, which cover the inputs and outputs as well as additional processing parameters and the underlying algorithm for each process. Within the proposed system a WPS instance is the main entry point to all predefined scripts and custom tools that are deployed within a cloud platform, which solely are executable from within the cloud (Figure \ref{fig:wps_processing}). The WPS instance itself defines all processes it supports and can be executed by any requesting client. For each provided process, a corresponding tool exists within the cloud environment, which will be triggered by the WPS. It is up to each WPS process to implement the tool invocation. Thus, the WPS interface hides tool execution details from the calling client. As a result, the underlying cloud platform can be switched with ease since process calling clients only interact with the respective WPS instance that wraps the tools within a certain processing environment.

Assuming, the underlying cloud platform provides a runtime environment for containers, each EO tool is encapsulated as Docker container. This facilitates the portability of containerized applications across different platforms. In addition, it is conceivable to deploy and run EO tools on several virtual machine instances if there is an increased need for computing resources. Therefore, horizontal scalability of EO processes on-demand is supported.

In conclusion, the WPS approach provides a standardized way to the remaining WaCoDiS system for automated and scalable EO processing in the cloud. Thus, the proposed approach supports the portability of EO processing services and prevents a vendor lock-in.

\subsection{Provision of analysis results for water monitoring tasks}
\label{Provision of analysis results for water monitoring tasks}
Conventional Spatial Data Infrastructures (SDIs) often lack the provision of up-to-date data without any manual interaction since they rely on request/response patterns \citep{rieke_geospatial_2018}. Thus, implementing a full automated processing pipeline, including the dissemination of processing results, is hard to achieve if it comes to the integration of existing SDIs. The WaCoDiS System addresses this issue by a component that automatically ingests newly generated products into existing data infrastructures. 

\begin{figure}[b]
    \centering
    \includegraphics[width=12cm]{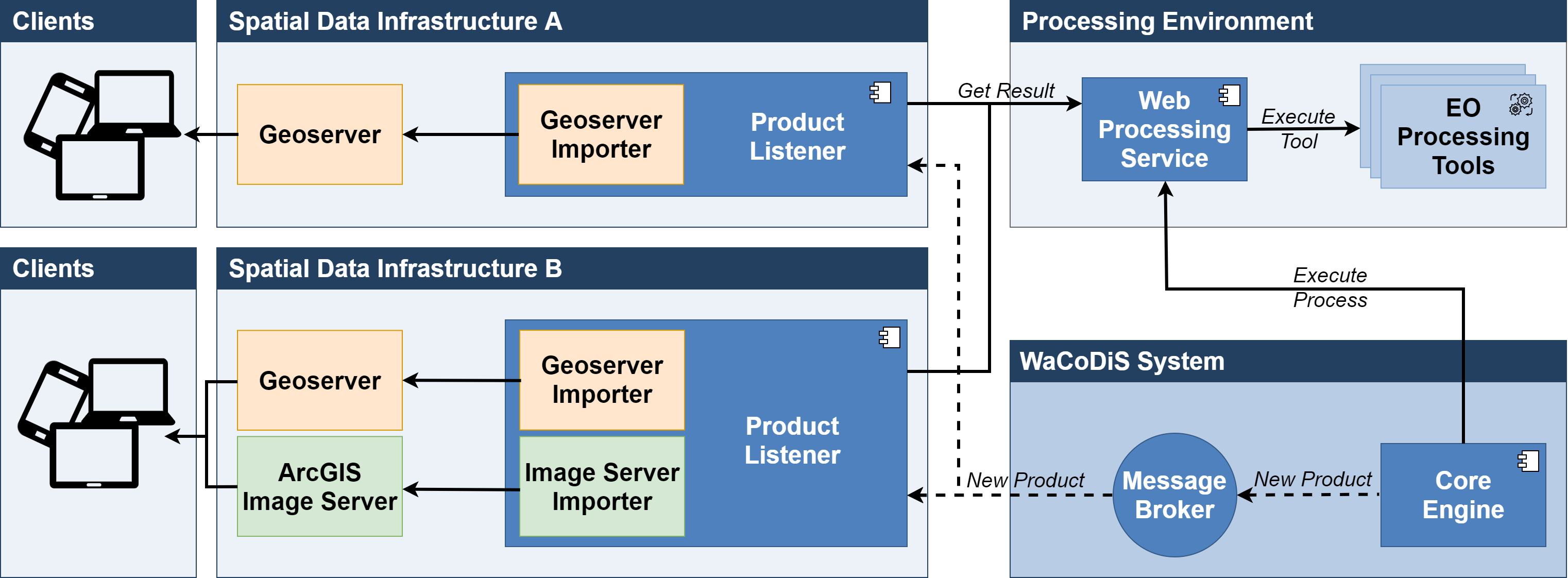}
    \caption{Overview of the components and message flows involved in data ingestion}
    \label{fig:data_ingestion}
\end{figure}

The Product Listener component of the system architecture (see \ref{System components for processing tasks}) implements the data ingestion. Aligned with the event-driven approach of the system architecture, the Product Listener is notified as soon as a processing job is completed and an EO-product has been successfully generated. The notification message includes a reference (hyperlink) to download the processed EO-product and additional metadata. The Product Listener downloads the referenced product from the processing environment and ingests it into one or more specified data backends. Connectivity to the various backends is achieved by implementing a common interface. The current implementation offers capabilities to ingest products into instances of GeoServer or ArcGIS Image Server. Connectivity for further backend types can be added by implementing the specified interface. Note, that a single Product Listener instance must run in an environment that allows access to both, the processing environment and the configured data backend but is capable of serving multiple backends in the same environment.

Figure \ref{fig:data_ingestion} illustrates the ingestion of EO-products into different spatial data infrastructures. The Core Engine initiates a processing job by calling the processing environment’s WPS interface (\textit{Execute Process}). The WPS server then executes the appropriate processing tool (\textit{Execute Tool}). After successful process execution, the Core Engine notifies all Product Listener instances about the availability of a new product by publishing a message (New Product) via the Message Broker. The Product Listeners retrieves the new product from the WPS instance (\textit{Get Result}) and finally invokes the appropriate interface implementation (Importer) to store it in a specific backend.

The Product Listener offers the capability to compile metadata for generated EO-products. The metadata is then indexed by the Data Wrapper component of the WaCoDiS system. This means that products themselves can be input data for other, higher-value products.

\section{System validation}
In section 4 we identified challenges regarding the facilitation of Big Earth Observation data processing. Different requirements considering event-based data flows as well as standardized and platform independent processing interfaces have been postulated, which aim to enable an automated processing of EO data and to facilitate the exploitation of EO products by domain related users. In order to validate our proposed system design against the requirements, we tested our system within a pre-operational deployment that sets up on the novel CODE-DE platform (section \ref{Interoperable EO data processing on CODE-DE}) and interconnects to the existing Spatial Data Infrastructure of the Wupperverband (section \ref{Ingestion of EO products within a domain related SDI}). 

\begin{figure}[htp]
    \centering
    \includegraphics[width=12cm]{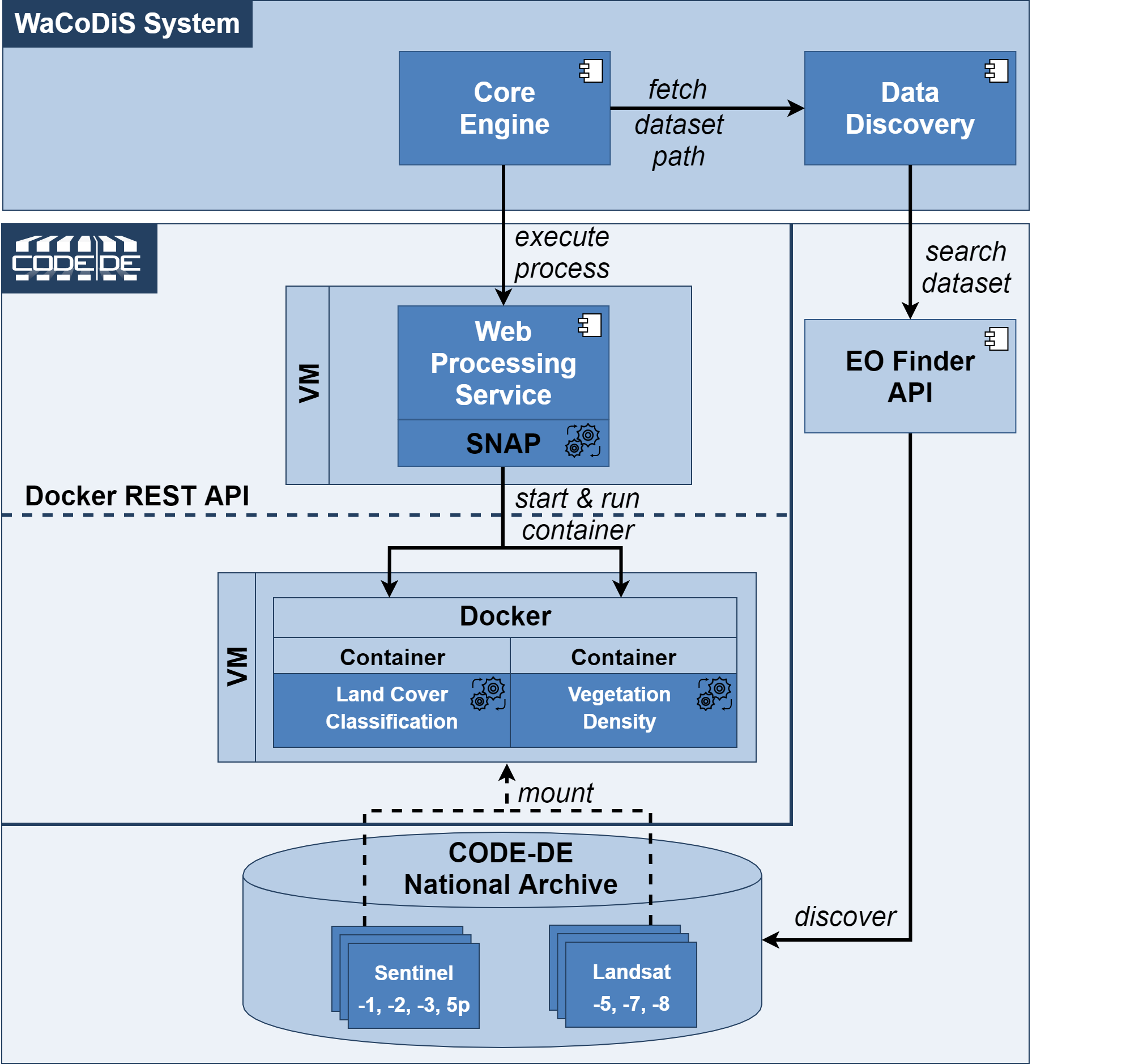}
    \caption{Architectural approach for encapsulating containerized Earth Observation processes by an OGC Web Processing Service within the CODE-DE cloud}
    \label{fig:code_de_deployment}
\end{figure}

\subsection{Interoperable EO data processing on CODE-DE}
\label{Interoperable EO data processing on CODE-DE}
The CODE-DE platform provides free access to datasets of the European Copernicus EO-program and, since its relaunch in April 2020, to a comprehensive processing environment. Capabilities of the CODE-DE cloud can be used via virtual machines (VM), which have direct access to Sentinel and Landsat collections. Computation usage is only limited by means of the amount of virtual computing resources that can be allocated by a user. Each virtual machine in the cloud will be started with a mounted directory, that contains all the satellite data collections from the CODE-DE National Archive. Thus, satellite datasets can be consumed directly within any arbitrary software (e.g., GIS, Sentinel Toolbox, Python scripts) without downloading the datasets. 

Our pre-operational setup comprised two virtual machines: one for EO data processing and another one for hosting the WPS (Figure \ref{fig:code_de_deployment}). The first VM has a public IP and hosts the WPS which enables the execution of the aforementioned tools from outside the cloud via its standardized interface. The second VM aims to run custom developed algorithms for generating desired EO products. These algorithms are containerized with Docker and can be started on demand from within the cloud. 

Our setup demonstrated the interoperability of the WPS approach, which enables a seamless integration of different EO processes into the automated processing flows covered by the WaCoDiS System. There is no need to connect separately to each EO tool, which runs on the CODE-DE cloud, since the WPS specification provides a standardized interface for process execution. Hence, new developed EO tools can be introduced to the system in a lightweight way, without the need to adjust the processing workflow within the WaCoDiS System.

Furthermore, we found that an important aspect is the provision of references on satellite datasets, which are discovered and registered by the Data Discovery components of the WaCoDiS system (in particular, Datasource Observer and Data Wrapper) via CODE-DE’s EO Finder API, to the WPS, instead of the actual data. Resolving the references and accessing the actual data to execute the tools will be done only from within the CODE-DE cloud. This pattern facilitates efficient processing near the data’s physical storage location.

However, some shortcomings regarding the EO tool execution have been detected. Since the EO tools are containerized but do not run as autonomous services, the WPS instance needs to know the VM which hosts a certain tool in beforehand. Then, starting the respective container and executing the processing tool can be triggered by the WPS server over the Docker REST API. Thus, the proposed approach not fully matches the „Software as a Service” paradigm.

\subsection{Ingestion of EO products within a domain related SDI}
\label{Ingestion of EO products within a domain related SDI}
\begin{figure}[t]
    \centering
    \includegraphics[width=12cm]{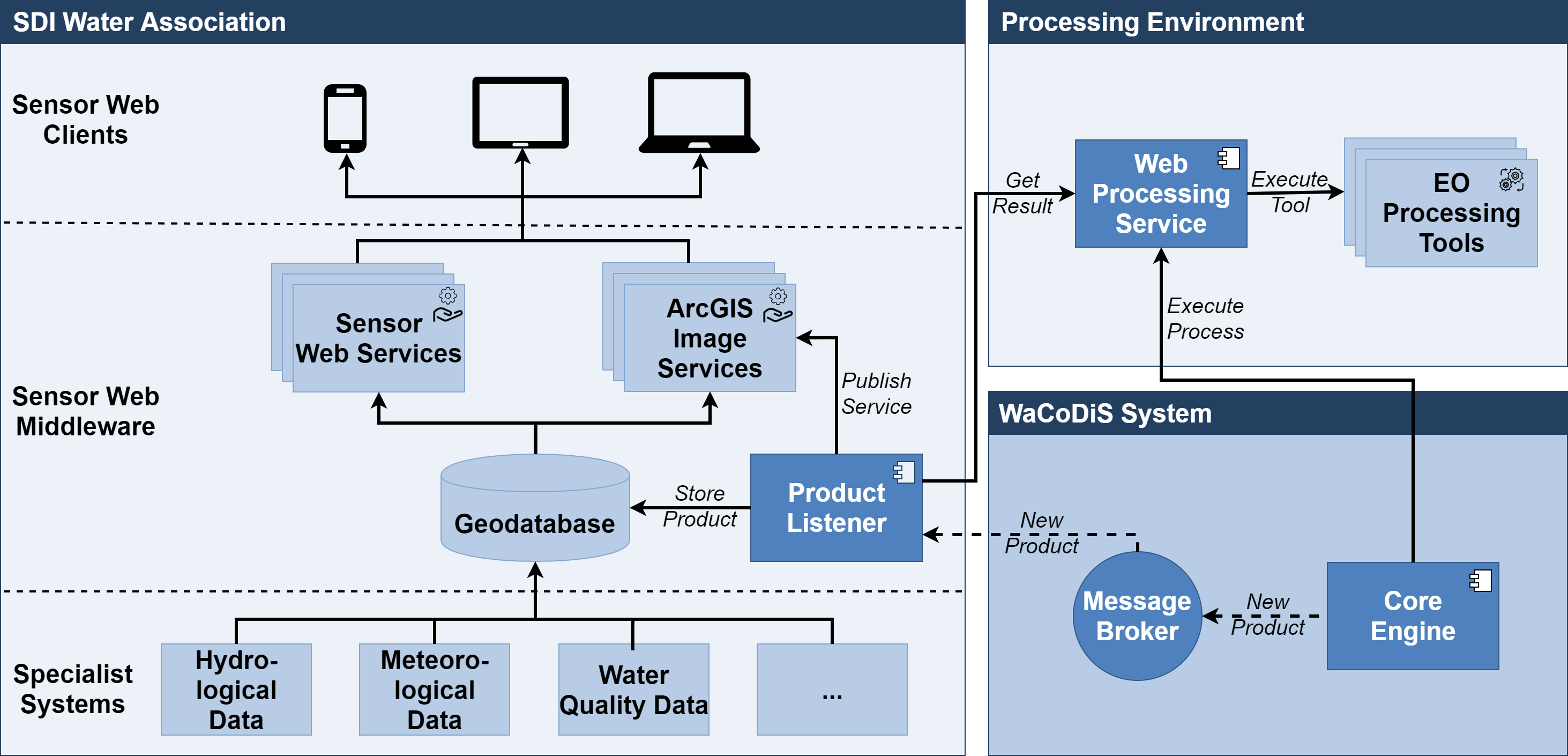}
    \caption{Event-driven ingestion of Earth Observation products within the SDI of a water management association via WaCoDiS Product Listener component}
    \label{fig:data_ingestion_wv_gdi}
\end{figure}

As part of the WaCoDiS project, the developed system architecture aimed to support different specialist departments of the Wupperverband for typical water management and monitoring tasks. Thus, EO products with valuable impact on water monitoring had to be disseminated within the Wupperverband’s SDI, which comprises several expert systems as well as a Sensor Web infrastructure for in-situ measurement data. This is implemented by deploying the proposed Product Listener approach (section \ref{Provision of analysis results for water monitoring tasks}) within the Wupperverband’s internal IT infrastructure (Figure \ref{fig:data_ingestion_wv_gdi}).

WaCoDiS Product Listener receives notifications about successfully finished remote sensing processes, calls up the product result via the WPS instance and makes it available in the system environment. For this purpose, the Product Listener implements an interface for the publication of Image Services in an ArcGIS Image Server. Finally, remote sensing products are made available in expert systems and web-based applications utilized by the specialist departments to fulfill their monitoring tasks.

It's worth mentioning, that the proposed ingestion approach is not limited to the Spatial Data Infrastructure of the Wupperverband but can also be transferred to other system environments. Any backend embedded within a domain related SDI can be connected to the WaCoDiS system by implementing the Product Listener interface with a backend-specific ingestion routine. The Product Listener itself is loosely coupled with the remaining WaCoDiS System since it only has to listen for certain messages indicating the completion of a process execution. However, the Product Listener does not provide a standardized interface for EO product publication. Thus, it has to implement the ingestion routine for each new backend type rather than interconnecting a certain backend directly to the WaCoDiS System via interoperable Publish/Subscribe interfaces for data publication.

\section{Conclusion}
Our paper introduced an event-driven approach for automated processing of EO data. The underlying architecture heavily relies on a Publish/Subscribe messaging pattern, using a message broker that serves as middleware between the system’s components. It was shown that the event driven approach is suitable for efficient EO data processing. All subtasks of a comprehensive processing pipeline, including data discovery, acquisition, processing and ingestion of generated results, are reflected in the software architecture. The individual subtasks are logically separated from each other and implemented separately. The message flow connects the resulting system components.

Most legacy data platforms and download services do not support the publish-subscribe pattern. In order to integrate data sources that provide EO data and additional required input data it was necessary to connect these sources by using a more traditional request-response pattern. Recent developments such as the OGC Sensor Things API promise a seamless integration of measurement data into event-driven system architectures. However, it remains to be seen whether this or similar developments will be implemented more frequently in the future for EO data.

Up to now, the current system triggers recurring processing workflows based on a configurable schedule, the workflow is finally executed when all required input data is available. A possible but outstanding extension of the described system would be the event-based execution of EO processing jobs. For example, the processing workflow could be triggered if an extreme weather event is detected in sensor-data streams. Such methods for real-time event detection have been addressed by recent research projects, such as SenSituMon\footnote{\url{https://www.sensitumon.eu/en/}}.

The WaCoDiS System comprises components for ingesting resulting EO products into one more data backends embedded in different environments. Thus, our approach extends existing domain related Spatial Data Infrastructures by Publish/Subscribe mechanisms for new data products. However, the implemented approach relies on a custom interface using mainstream technologies since geospatial IT standards neglect event-driven data flows. Consequently, interoperability regarding near real-time data streams is still an outstanding task within the geospatial world.

By relying on standardized interfaces, the interoperability of the automated workflow for processing EO data is promoted. The OGC WPS standard provides a uniform interface that encapsulates different processes for generating higher-value EO products. The system components can be deployed in various environments (e.g., cloud platforms, self-hosted servers). Hence, it is possible to execute the algorithms close to the Big EO data and avoid time-consuming data transfers. However, just pure cloud platform providers such as Amazon offer both, access to Sentinel imagery and tools for service orchestration (e.g., Kubernetes). CODE-DE provides seamless access to the data acquired as part of the European Copernicus program but has neither a ready-to-use service orchestration that is required to deploy the WaCoDiS core system’s components nor a built-in messaging system. 

Furthermore, other datasets than EO data (e.g., in-situ measurement data or domain related datasets) are not available on the same cloud platform and still need to be transferred to the processing environment. The resulting disadvantage is manageable since the EO data usually have a much larger volume than the additional input data, but a completely cloud-native processing is therefore subject of future work.

\section{Acknowledgements}
This work has been co-funded as part of the WaCoDiS project by the Federal Ministry of Transport and Digital Infrastructure (Germany), BMVI, as part of the mFund program.

\bibliographystyle{apalike}
\bibliography{references} 

\end{document}